%% file: EVC-PT.tex
\documentclass[aps,prl,twocolumn,superscriptaddress,preprintnumbers,showkeys,nofootinbib]{revtex4-1}

\usepackage[utf8]{inputenc}
\usepackage[english]{babel}
\usepackage{amssymb,amsthm,amsmath,amstext,amsbsy,amsopn}
\usepackage{bbm}
\usepackage{nicefrac}
\usepackage{slashed}
\usepackage{graphicx}
\usepackage{hyperref}
\usepackage{leftidx}
\usepackage{environ}
\usepackage{mathtools}
\usepackage{xspace}
\usepackage{array}
\usepackage{xcolor}
\usepackage{pgfplotstable}
\usepackage{isotope}

\usepackage[normalem]{ulem}

\hyphenchar\font=\string"7F
\input{defs}

\begin{document}

\title{Improved many-body expansions from eigenvector continuation}

\author{P.~Demol}
\affiliation{KU Leuven, Instituut voor Kern- en Stralingsfysica, 3001 Leuven,
Belgium}

\author{T.~Duguet}
\affiliation{KU Leuven, Instituut voor Kern- en Stralingsfysica, 3001 Leuven,
Belgium}
\affiliation{IRFU, CEA, Universit\'e Paris-Saclay, 91191 Gif-sur-Yvette, France}

\author{A.~Ekstr\"om}
\affiliation{Department of Physics, Chalmers University of Technology, 412
96 G\"oteborg, Sweden}

\author{M.~Frosini}
\affiliation{IRFU, CEA, Universit\'e Paris-Saclay, 91191 Gif-sur-Yvette, France}

\author{K.~Hebeler}
\affiliation{Institut f\"ur Kernphysik, Technische Universit\"at Darmstadt,
64289 Darmstadt, Germany}
\affiliation{ExtreMe Matter Institute EMMI,
GSI Helmholtzzentrum f\"ur Schwerionenforschung GmbH,
64291 Darmstadt, Germany}

\author{S.~K\"onig}
\affiliation{Institut f\"ur Kernphysik, Technische Universit\"at Darmstadt,
64289 Darmstadt, Germany}
\affiliation{ExtreMe Matter Institute EMMI,
GSI Helmholtzzentrum f\"ur Schwerionenforschung GmbH,
64291 Darmstadt, Germany}

\author{D.~Lee}
\affiliation{National Superconducting Cyclotron Laboratory,
Michigan State University, MI 48824, USA}


\author{A.~Schwenk}
\affiliation{Institut f\"ur Kernphysik, Technische Universit\"at Darmstadt,
64289 Darmstadt, Germany}
\affiliation{ExtreMe Matter Institute EMMI,
GSI Helmholtzzentrum f\"ur Schwerionenforschung GmbH,
64291 Darmstadt, Germany}
\affiliation{Max-Planck-Institut f\"ur Kernphysik,
Saupfercheckweg 1,
69117 Heidelberg, Germany}

\author{V.~Som\`a}
\affiliation{IRFU, CEA, Universit\'e Paris-Saclay, 91191 Gif-sur-Yvette, France}

\author{A.~Tichai}
\email{alexander.tichai@physik.tu-darmstadt.de}
\affiliation{ESNT, CEA, Irfu, D\'epartement de Physique Nucl\'eaire,
Universit\'e Paris-Saclay, F-91191 Gif-sur-Yvette}
\affiliation{Max-Planck-Institut f\"ur Kernphysik,
Saupfercheckweg 1,
69117 Heidelberg, Germany}
\affiliation{Institut f\"ur Kernphysik, Technische Universit\"at Darmstadt,
64289 Darmstadt, Germany}
\affiliation{ExtreMe Matter Institute EMMI,
GSI Helmholtzzentrum f\"ur Schwerionenforschung GmbH,
64291 Darmstadt, Germany}


\begin{abstract}
Quantum many-body theory has witnessed tremendous progress in various fields, 
ranging from atomic and solid-state physics to quantum chemistry and
nuclear structure.  Due to the inherent computational burden linked to the
\abinitio treatment of microscopic fermionic systems, it is desirable to obtain
accurate results through low-order perturbation theory. In atomic nuclei however, 
effects such as strong short-range repulsion between 
nucleons can spoil the convergence of the expansion and make the reliability of 
perturbation theory unclear. Mathematicians have devised an extensive
machinery to overcome the problem of divergent expansions by making use of
so-called resummation methods.  In large-scale many-body applications such
schemes are often of limited use since no \apriori analytical knowledge of the
expansion is available.  We present here eigenvector continuation as an
alternative resummation tool that is both efficient and reliable because it
is based on robust and simple mathematical principles.
\end{abstract}

\maketitle

\paragraph{Introduction}

The quantum-mechanical treatment of many interacting particles from first
principles poses a formidable formal and computational challenge.  At the heart
of the \abinitio philosophy is the capacity to control the error induced by
many-body approximations used to solve the stationary Schr\"odinger equation.
Light systems, e.g., atomic nuclei with small mass number $A$, can be accurately
described by large-scale diagonalization approaches like configuration
interaction (CI)~\cite{NaQu09,RoLa11,BaNa13} or by imaginary-time propagation
methods via Quantum Monte Carlo (QMC)
techniques~\cite{Gezerlis2013,Carlson:2015,Lynn:2017}.  However, the exponential
growth of the Hilbert-space dimension with increasing $A$ constitutes an ordeal
for many-body practitioners interested in systems containing tens or hundreds of
particles. In this Letter we present a novel approach to the many-body problem
that uses the method of eigenvector continuation to accelerate the convergence
of calculations using many-body perturbation theory. As we show, the method can 
produce accurate results even in cases where standard approaches fail to converge.

Expansion methods are a powerful tool for \abinitio calculations.  The exact
eigenstate of the Hamiltonian $H$ is systematically expanded around a simple,
yet appropriately chosen, $A$-body reference state.  Truncating the expansion to
a given order translates into a polynomial scaling with system size, making the
calculation possible for up to 100 fermions~\cite{Morris:2018}.  Most
commonly used approaches of this kind are many-body perturbation theory
(MBPT)~\cite{Bogner:2005,Roth:2009up,Hebeler:2011,Langhammer2012,Holt:2014,Tichai2016,Hu:2016,Tichai:2018ncsmpt,Tichai:2018mll,Drischler:2019},
coupled-cluster (CC) theory~\cite{HaPa10,BiLa14,Henderson:2014vka,Hagen:2014review},
self-consistent Green's function (SCGF)
theory~\cite{Dickhoff:2004xx,CiBa13,Carbone:2013eqa,SoCi13,Raimondi:2019} and
the in-medium similarity renormalization group
(IMSRG)~\cite{Tsukiyama:2011,Tsukiyama:2012,HeBo13,Bo14,H15,Parzuchowski2017,Stroberg2017}.
While they have been used with great success in various fields of many-body
research for a long time, the (re-)import into nuclear physics of
CC~\cite{KoDe04} and SCGF~\cite{Barbieri2002} from quantum chemistry about 15
years ago played a decisive role to reestablish \abinitio nuclear many-body
theory as a viable route to study nuclei with more than ten nucleons.  Due to the
strong short-range repulsion that is a feature of many, although not all,
representations of nuclear forces, it is only much more recently that MBPT has
been reconsidered as a viable
option~\cite{Bogner:2005,Holt:2014,Tichai2016,Hu:2016,Tichai:2018ncsmpt,Tichai:2018mll,Drischler:2019}.

In fact, there are two characteristics of the two-nucleon interaction making the
many-body problem hard to solve, e.g., \apriori non-perturbative~\cite{Bogner:2005,Bogner:2006,Ramanan:2007,Hoppe:2017}.
The first one relates to strong high-momentum correlations induced by the short-range repulsive interactions noted above.  
Fortunately, this problem can be tamed via the use of renormalization group transformations. 
The second issue stems from the large scattering lengths corresponding with resonant scattering in the S-wave channels, 
which can produce strong many-body correlations.  While Pauli-blocking and effective-range effects ameliorate 
this problem to some extent, one of the motivations of approaches such as Bogoliubov many-body perturbation theory (BMBPT) 
is that the use of symmetry-breaking reference states allows one to include more long-range correlations into the reference state, 
which can help to tame this problem further in nuclei.

While we have reduced some of the problems of strong correlations, we may find that the perturbative 
expansion is still not convergent and requires using resummation tools.  However, since the analytical properties 
of the many-body expansion are usually unknown, conventional resummation methods often cannot be applied with confidence.
Therefore, an alternative framework that does not require such a knowledge is
highly desirable. In this Letter, we present eigenvector continuation (EC)
as a technique that allows to achieve this goal. Two
numerical applications dedicated to the closed- and open-shell nuclei $^{3}$H
and $^{18}$O, respectively, are provided to illustrate the power of the method.

\paragraph{Eigenvector continuation}
\label{sec:EC}

Recently, the EC technique was introduced~\cite{Frame:2018,Koenig:2019} to treat physical
systems the Hamiltonian of which is a particular instance, e.g., $H\equiv H(1)$,
of an operator $H(c)$ depending smoothly on a parameter $c$.  The power of EC
relies on
\begin{enumerate}
\item the fact that there exists a regime, $0\leq c \leq c_e < 1$, for
 which the many-body problem is easier to solve than for the target value
 $c=1$,
\item the stability of the eigenvectors of $H(c)$ against variations of $c$,
 i.e., the targeted many-body state remains in a low-dimensional manifold of the
 full $A$-body Hilbert space when changing $c$ from $[0,c_e]$ to $1$.
\end{enumerate}
Based on these principles, and targeting a particular eigenstate of $H$, e.g.,
the nuclear ground state, EC works in two successive steps
\begin{enumerate}
\item a low-dimensional manifold of $\NEC$ auxiliary states $\{\ket{\Psi(c_i)};
 i=1,\ldots\NEC\}$ is obtained by solving, to the best of one's capacity, the
 $A$-body Schr\"odinger equation associated with $H(c_i)$, $c_i \in [0,c_e]$,
\item the physical Hamiltonian $H(1)$ is diagonalized within the low-dimensional
 manifold obtained in step 1.  The auxiliary states being non-orthogonal, solving
 the secular equation requires the computation of two $\NEC\times\NEC$ matrices,
 i.e., the Hamiltonian kernel $H_{ij} \equiv
 \mbraket{\Psi(c_i)}{H(1)}{\Psi(c_j)}$ and the norm kernel $N_{ij} \equiv
 \braket{\Psi(c_i)}{\Psi(c_j)}$.
\end{enumerate}
In practice, it is typically advantageous to first diagonalize the norm matrix and
eliminate eigenvectors associated with its very small eigenvalues that arise
when choosing the $c_i$ from a narrow range.

\paragraph{Perturbation theory}

The EC method is in principle agnostic with respect to the particular
computational method used to generate the $\NEC$ auxiliary states
$\{\ket{\Psi(c_i)}\}$.  In practice, the performance of EC does of course
depend on the ability to describe the ground state of $H(c_i)$ with sufficient
accuracy.  In this Letter, the low-dimensional EC manifold is built from
perturbative corrections on top of a well-chosen reference state.  While
the perturbative expansion is in general not suited to reach directly $H(1)$, EC
can be understood as effectively performing a \emph{sequence} of analytic
continuations and is thus able to go beyond the radius of convergence of the
perturbative expansion while using the same inputs.  It has been observed that
EC even works in cases where the radius of convergence of perturbation theory is
strictly zero~\cite{Sarkar:2019}.  The only requirement is that the dependence
of $H(c)$ on $c$ is sufficiently smooth.

The procedure starts from the partitioning of the Hamiltonian according
to $H \equiv H_0 + H_1$ such that the reference state is the ground
state of the unperturbed Hamiltonian
\begin{align}
 H_0 |\Phi^{(0)}\rangle = E^{(0)} |\Phi^{(0)}\rangle \, .
\end{align}
The eigenstates of $H_0$ obtained through elementary, e.g., particle-hole,
excitations of $|\Phi^{(0)}\rangle$ provide an orthonormal basis of the
many-body Hilbert space.  Scaling the residual interaction $H_1$ by a parameter
$c$ to introduce the parameter-dependent Hamiltonian $H(c) \equiv H_0 + c \,
H_1$, perturbation theory (PT) generically parametrizes the exact ground state
of the latter via an infinite power series
\begin{align}
 |\Psi (c) \rangle \equiv \sum_{p=0}^\infty c^p |\Phi^{(p)} \rangle \, ,
\label{MBPTstate}
\end{align}
characterized by an (unknown) radius of convergence $c \in [0, R_c]$.  In
Eq.~\eqref{MBPTstate}, $|\Phi^{(p)}\rangle$, $p \geq 1$, denotes the
perturbative state correction of order $p$, which is independent of $c$ and
typically given as a specific linear combination of the eigenstates of
$H_0$~\cite{Shavitt2009}.

Choosing $0 \leq c_i \leq c_e \leq R_c$, $i=1,\ldots\NEC$ and working at PT
order $P$, the approximate low-dimensional manifold can be related to the
reference state and the first $P$ state corrections via the transformation
\begin{align}
\begin{pmatrix}
 | \Psi_{P} (c_1) \rangle \\
 | \Psi_{P} (c_2) \rangle \\
 \vdots \\
 | \Psi_{P} (c_{\NEC}) \rangle
 \end{pmatrix}
 =
 \begin{pmatrix}
 1      &c_1            &c_1^2  &\cdots &c_1^P \\
 1      &c_2            &c_2^2  &\cdots &c_2^P \\
 \vdots & \vdots        &\vdots &\ddots &\vdots \\
 1      &c_{\NEC}       &c_{\NEC}^2     &\cdots &c_{\NEC}^P
 \end{pmatrix}
 \begin{pmatrix}
 | \Phi^{(0)} \rangle \\
 | \Phi^{(1)} \rangle \\
 \vdots \\
 | \Phi^{(P)} \rangle
 \end{pmatrix}
 \nonumber
\end{align}
such that the EC eigenvalue problem is equivalently formulated within
the manifold $\{| \Phi^{(p)} \rangle; p=1,\ldots P\}$.  Eventually, the actual
values $\{c_i ; i=1,\ldots\NEC\}$ do not matter and the dimensionality of the
manifold is effectively set by the perturbative order $P$.  One thus needs to
compute the $(P+1)\times (P+1)$ matrices
\begin{subequations}
\begin{align}
 \mathbf{H}_{pq} &\equiv \langle \Phi^{(p)} | H | \Phi^{(q)} \rangle \, ,\\
 \mathbf{N}_{pq} &\equiv \langle \Phi^{(p)} |  \Phi^{(q)} \rangle \, ,
\end{align}
\end{subequations}
and solve the secular equation
\begin{align}
 \mathbf{H} \mathbf{N} X = E \mathbf{N} X \, , \label{secular}
\end{align}
where the diagonal matrix $E$ gathers $P+1$ eigenenergies, the lowest of which
relates to the ground state.\footnote{Although the ground state is targeted,
Eq.~\eqref{secular} actually delivers $P+1$ vectors and energies.  It will be
interesting to investigate to what extent the excited states within the EC
subspace can be associated with states in the spectrum of $H$ carrying the
same quantum numbers as the ground state.} Eventually, the procedure reduces to
calculating PT state corrections at a chosen order $P$, computing the
$(P+1)\times (P+1)$ Hamiltonian and norm matrices as well as solving the
associated secular equation.  Last but not least, the eigenvectors can further
be used to compute other observables of interest.

While the present Letter provides proof-of-concept results up to high orders
$P$, future work will target fully realistic calculations at low orders,
e.g., $P=3$, that are currently within reach of state-of-the-art numerical
codes implemented in large model spaces.

\paragraph{Applications}

We now consider the MBPT-based EC scheme applied to two situations of increasing
complexity, with the goal of illustrating the difficulties posed by the two
sources of non-perturbative behavior mentioned at the outset.  In both cases, the calculations are
performed up to high perturbative orders based on a recursive scheme and the
interaction employed is the EM500 interaction~\cite{Entem2003a}, with
three-nucleon forces omitted for simplicity. Furthermore, a similarity
renormalization group (SRG) transformation characterized by a continuous flow
parameter $\lambda$ is applied to soften the nuclear
Hamiltonian~\cite{Bogner:2007}.
 
\begin{figure}[t!]
\centering
\includegraphics[width=1.0\columnwidth]{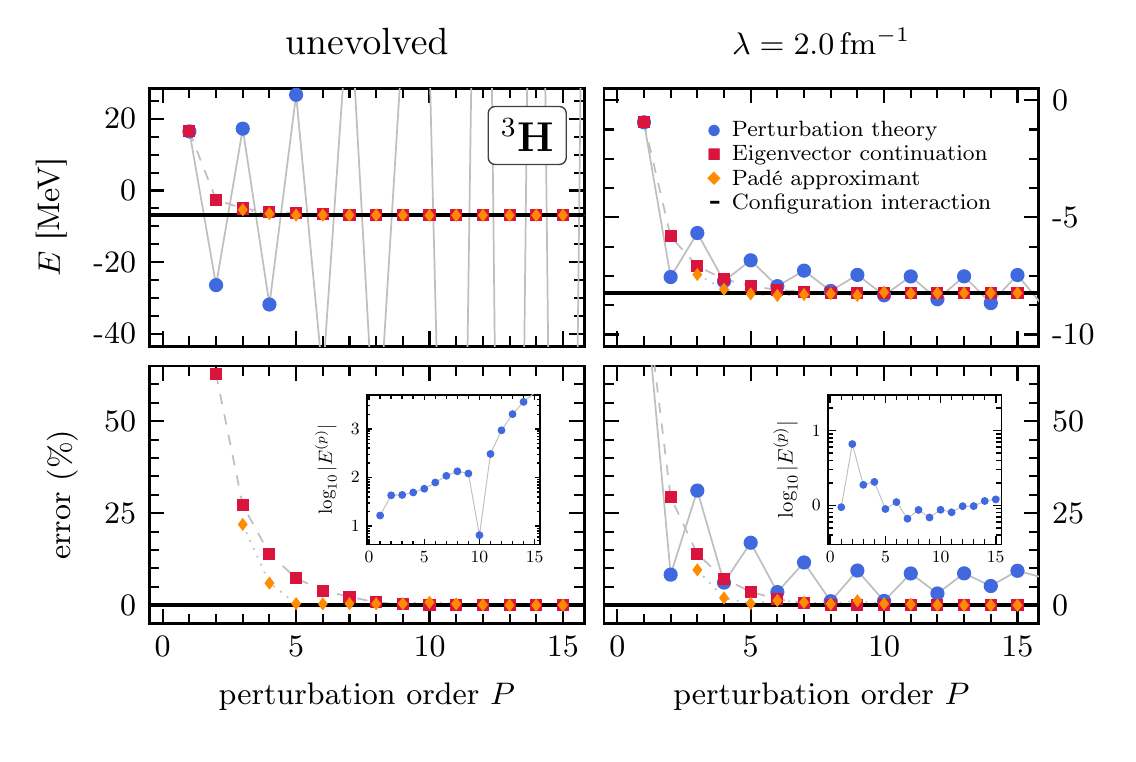}
\caption{%
Ground-state energy of $^{3}$H for MBPT (blue circles), MBPT-based-EC (red
squares) and MBPT-based-Padé (yellow diamonds) as a function of the perturbative order $P$ against exact CI diagonalization
(full line).  Left panels: unevolved Hamiltonian.  Right panels:  $\lambda = 2.0$ fm$^{-1}$.  Top panels: absolute
energies. Bottom panels: error relative to the CI result. The insets display the successive perturbative contributions in logarithmic scale.}
\label{fig:jacobi}
\end{figure}

The first application is dedicated to the very light nucleus $^{3}$H.  The exact
result and the recursive scheme to perform MBPT are based on a CI code built
within the harmonic-oscillator Jacobi-coordinate NCSM
formalism~\cite{Navratil:1999pw}.  The large three-body Hilbert (sub)space employed
includes configurations up to $N_{\text{max}} = 12$ excitations above the
harmonic-oscillator Slater-determinant reference state. In Fig.~\ref{fig:jacobi}, the 
ground-state energy is displayed as a function of the MBPT order $P$ for the unevolved EM500 interactions and for the SRG-evolved potential with $\lambda=2.0\,\text{fm}^{-1}$.  Results from the CI diagonalization are compared 
to MBPT as well as to EC and (diagonal and super-diagonal) Pad\'e resummations built on top of it~\cite{Baker:1996}.

When employing a ``hard'' interaction displaying large low-to-high momentum coupling
(left panels), the perturbative series exhibits a divergent behaviour.
Using EC on top of it yields a rapid and monotonic convergence towards the exact CI result. 
The EC and Pad\'e results are very similar. The effect of ``softening'' the short-range behavior of the
interaction via the SRG transformation is seen in the right panels: the
perturbation series oscillates much more mildly and seems to converge towards the CI
limit, although it in fact diverges at even higher orders.  Even in this case EC improves upon MBPT by exhibiting a monotonic and
rapid convergence providing, at each order, a variational upper bound to
the CI result. Again, EC and Pad\'e results are very similar. At $P=4$, their deviation from the CI result is already below $5\%$. The same behavior is found for \isotope[4]{He} and for various SRG
parameters.

Eigenvector continuation based on the first few MBPT orders manages to reveal
the exact result even for the unevolved Hamiltonian, where the series
displays a divergence of ultraviolet character.  It must be noted that, while
the above calculations have been performed using a reference Slater determinant
built from the spherical harmonic-oscillator one-body basis, the divergence of
the MBPT series can actually be overcome by starting from an optimized (e.g.,
Hartree-Fock) mean-field reference state~\cite{Tichai2016}.  In open-shell
nuclei, the infrared divergence associated with the emergence of nuclear
superfluidity and/or deformation requires an even more drastic optimization of
the reference state, i.e., of $H_0$.  This can be achieved by using symmetry-breaking 
references states, e.g., using a Bogoliubov extension of MBPT, as we will discuss in the following.


Neither EC nor any other resummation technique is useful when the reference state 
is degenerate with respect to elementary excitations since this prevents the construction of the first few perturbative orders.  A
powerful way to resolve the issue is to allow $H_0$ and $| \Phi^{(0)} \rangle$
to break symmetries of $H$.  For example, nuclear superfluidity can be handled at
the price of breaking $U(1)$ symmetry associated with the conservation of
particle number.  In this case, the reference state becomes a Bogoliubov vacuum
that already captures (most of) the so-called static correlations originally
responsible for the infrared divergence.  Building perturbation theory on top of
such a reference state led recently to the introduction of BMBPT~\cite{Duguet:2015yle}.  In BMBPT, the average
particle number must be monitored and adjusted to the targeted value $A$ at each
perturbative order.  Thus, BMBPT qualifies as a perturbation theory \emph{under
constraint}~\cite{demol19a}.  Recent large-scale second and third-order BMBPT
calculations of mid-mass open-shell nuclei demonstrated the merits of the method
and provided ground-state energies in agreement with the most sophisticated
non-perturbative many-body schemes on a few percents level at a small fraction
of the computational cost~\cite{Tichai:2018mll}.

The second application is thus dedicated to the open-shell nucleus $^{18}$O and
employs the soft interaction characterized by the SRG parameter
$\lambda=2.0\,\text{fm}^{-1}$.  Compared to $^{3}$H, the CI result and the
recursive scheme to perform BMBPT require a more drastic limitation of the model
space.  First, a one-body harmonic-oscillator basis utilized in the present
application is severely truncated\footnote{A converged \abinitio calculation
 with respect to the size of the one-body basis would typically require
$e_{\text{max}}=2n + l=13$~\cite{Tichai:2018mll}.} at $e_{\text{max}}=2n + l=4$.
A symmetry-broken Hartree-Fock-Bogoliubov (HFB) reference state is obtained in
that model space by solving the variational mean-field HFB
equations~\cite{RiSc80}.  Subsequently, the many-body basis of $H_0$ used to
expand perturbative state corrections is limited to configurations obtained via
two-, four- and selected~\cite{Tichai:2019ksh} six-quasiparticle excitations of
the reference state.\footnote{This corresponds to limiting the configuration
space to one-particle/one-hole, two-particle/two-hole and selected
three-particle/three-hole excitations when using a simpler Slater determinant
reference state.}  The latter truncation leads to performing approximate BMBPT
calculations at order $P\geq 3$ where basis states associated with eight
quasiparticles and beyond do contribute to state corrections.  The CI result
acting as a benchmark is obtained in the same configuration basis.

\begin{figure}
\centering
\includegraphics[width=1.0\columnwidth]{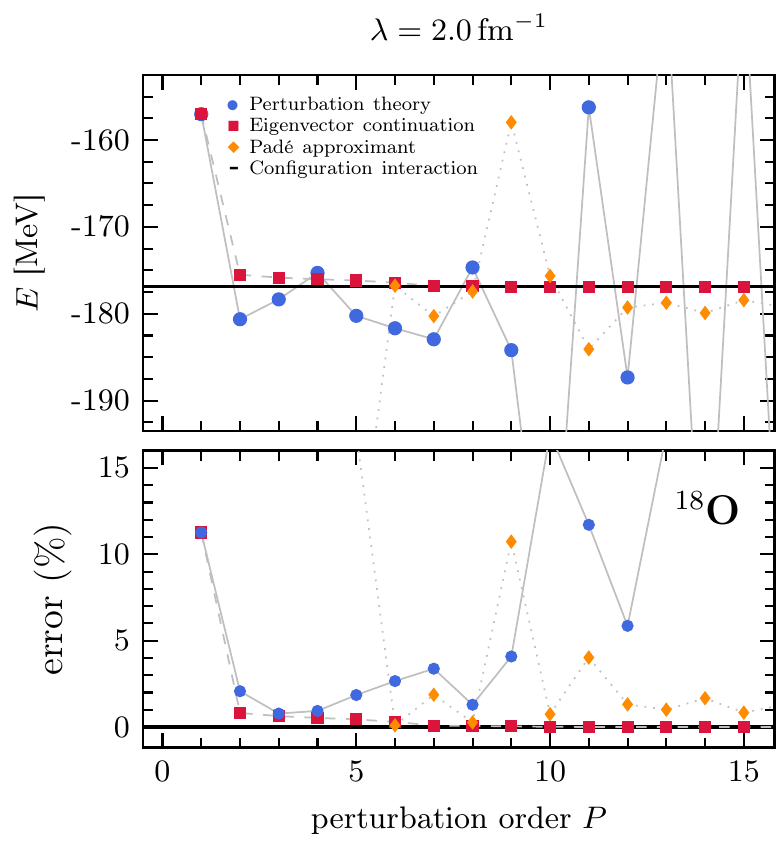}
\caption{%
Ground-state energy of $^{18}$O for BMBPT (blue circles), BMBPT-based-EC (red
squares) and BMBPT-based-Padé (yellow diamonds) as a function of the perturbative order $P$ against exact CI
diagonalization (full line) for $\lambda = 2.0$ fm$^{-1}$.  Top panel: absolute energies.  Bottom: relative error
to the CI result.
}
\label{fig:bmbpt}
\end{figure}

Figure~\ref{fig:bmbpt} displays the ground-state energy of $^{18}$O as a
function of the perturbative order $P$.  Results are shown for BMBPT as well as
for BMBPT-based (diagonal) Pad\'e resummation and EC.  One first observes that
variationally optimizing the $U(1)$-breaking Bogoliubov reference state does not
prevent the perturbative series to diverge.  Indeed, best coping with the most
dramatic, i.e., infrared, divergence is achieved at the price of inducing an
ultraviolet one~\cite{demol19a} that was not present in $^{3}$He for $\lambda =
2.0$ fm$^{-1}$.  Still, the tremendous benefit of the approach is to make low
MBPT orders meaningful in open-shell nuclei such that they actually provide a
decent account of the exact result.  Performing a parametric Pad\'e resummation
slightly reduces the oscillations for intermediate orders but the resummed
expansion still diverges.  Only with EC a rapid and monotonically converging
series towards the CI result is obtained.

As a matter of fact, EC based on a low-order BMBPT calculation provides an
accurate upper bound of the CI result, with sub-percent error for $P \geq 1$.
The method thus constitutes a very promising tool for fully
realistic\footnote{In addition to being converged with respect to the harmonic-oscillator
 one-body basis, realistic calculations refer here to results
performed with three-nucleon forces.} calculations where low-order BMBPT can be
applied at moderate computational cost.  One must of course beware that, by
continuing approximate eigenvectors $| \Psi_{P}(c) \rangle$ obtained at small
values of $c$, EC can at best account for the CI result at $c=1$ \emph{within
the eigensubspace} of $H_0$ covered by the approximate eigenvectors. Realistic
BMBPT(2) calculations probe at most zero, two, four and (selected) six (and in
principle eight) quasiparticle excitations.  Thus, the corresponding EC result
can capture non-perturbative correlations associated with up to six
quasi-particle (three-particle/three-hole) excitations, which is superior to
what available state-of-the-art non-perturbative methods can deliver today for
mid-mass (open-shell) nuclei.

\paragraph{Conclusions and Outlook}

In this Letter, a novel non-parametric resummation method was applied to
perturbative expansions appropriate to closed- and open-shell nuclei. Calculations were based on a realistic nuclear
two-body interaction derived from chiral effective field theory.  While the
perturbative expansion is divergent in most cases, EC
provides a robust framework to obtain a rapidly converging resummed sequence.

In particular, EC elegantly overcomes divergences arising from possible strong high-momentum couplings in nuclear interactions and/or induced by the interference between them and strong infrared correlations
as long as the perturbation theory exploits a symmetry breaking
reference state.  Benchmark calculations reveal that EC based on low-order
perturbation theory is vastly superior to parametric resummation schemes like
Pad\'e approximants that are widespread among many-body practitioners.
Consequently, EC provides an efficient and accurate tool to improve many-body
expansions independently of the origin of the Hamiltonian, thus being very
promising for various fields of many-body physics and chemistry.

While the results presented in this work are of proof-of-principle character, EC
will soon be applied to fully realistic \abinitio large-scale calculations of
mid-mass closed- and open-shell nuclei.  This will be done on top of low-order
perturbative calculations performed in a large single-particle basis set and
starting from realistic two- and three-nucleon interactions.  Since the EC
method yields a variational upper bound for binding energies, it also provides a
first step towards the quantification of many-body uncertainties from
non-variational many-body techniques.  Notably, EC also gives access to other
ground-state observables~\cite{Koenig:2019} and low-lying excitations at the same time.
Furthermore, the method can be applied to infinite nuclear matter calculations.

While the present work implements EC on top of perturbation theory, it can be
similarly applied on top of more sophisticated non-perturbative expansion
methods such as coupled cluster theory~\cite{Ekstroem:2019}.  It will be interesting to see if a
further optimization of the solution can be obtained in this way.  Be it
initially through perturbation theory or perturbatively corrected coupled
cluster, the fully non-perturbative treatment of three-particle/three-hole
excitations in mid-mass (and eventually heavy-mass) nuclei via EC  will be
superior to presently available implementations of many-body expansion methods.

\paragraph{Acknowledgments}

This work was supported by the European Research Council (ERC) under the European Union’s Horizon 2020 research and innovation programme (Grant agreement No. 758027), by the Deutsche Forschungsgemeinschaft (DFG, German Research Foundation) -- Projektnummer 279384907 -- SFB 1245, by the Espace de Structure et de r\'eactions
Nucl\'eaires Th\'eorique (ESNT) at CEA in France, and by the U.S.\ Department of Energy
(DE-SC0018638 and DE-AC52-06NA25396).  Some of the calculations were performed
by using HPC resources from GENCI-TGCC (Contract No. A005057392).

\bibliography{ECbib}

\end{document}

%% file: defs.tex
\newcommand{\apriori}{\textit{a priori}\xspace}

\newcommand{\abinitio}{\textit{ab initio}\xspace}

\newcommand{\ket}[1]{|#1\rangle}

\newcommand{\braket}[2]{\langle #1|#2\rangle}

\newcommand{\mbraket}[3]{\langle #1|#2|#3\rangle}

\newcommand*\rvec[1]%
{\ensuremath{\overset{\smash{\raisebox{-1.5pt}{\tiny$\rightarrow$}}}{#1}}}
\newcommand*\lvec[1]%
{\ensuremath{\overset{\smash{\raisebox{-1.5pt}{\tiny$\leftarrow$}}}{#1}}}

\NewEnviron{subalign}[1][]{%
\begin{subequations}\begin{align}
  \BODY
\end{align}\label{#1}\end{subequations}
}

\NewEnviron{spliteq}{%
\begin{equation}\begin{split}
  \BODY
\end{split}\end{equation}
}

\newcolumntype{K}[1]{>{\centering\arraybackslash}p{#1}}

\newcommand{\NEC}{\ensuremath{N_\text{EC}}}

%% file: EVC-PT.bbl
\begin{thebibliography}{51}%
\makeatletter
\providecommand \@ifxundefined [1]{%
 \@ifx{#1\undefined}
}%
\providecommand \@ifnum [1]{%
 \ifnum #1\expandafter \@firstoftwo
 \else \expandafter \@secondoftwo
 \fi
}%
\providecommand \@ifx [1]{%
 \ifx #1\expandafter \@firstoftwo
 \else \expandafter \@secondoftwo
 \fi
}%
\providecommand \natexlab [1]{#1}%
\providecommand \enquote  [1]{``#1''}%
\providecommand \bibnamefont  [1]{#1}%
\providecommand \bibfnamefont [1]{#1}%
\providecommand \citenamefont [1]{#1}%
\providecommand \href@noop [0]{\@secondoftwo}%
\providecommand \href [0]{\begingroup \@sanitize@url \@href}%
\providecommand \@href[1]{\@@startlink{#1}\@@href}%
\providecommand \@@href[1]{\endgroup#1\@@endlink}%
\providecommand \@sanitize@url [0]{\catcode `\\12\catcode `\$12\catcode
  `\&12\catcode `\#12\catcode `\^12\catcode `\_12\catcode `\%12\relax}%
\providecommand \@@startlink[1]{}%
\providecommand \@@endlink[0]{}%
\providecommand \url  [0]{\begingroup\@sanitize@url \@url }%
\providecommand \@url [1]{\endgroup\@href {#1}{\urlprefix }}%
\providecommand \urlprefix  [0]{URL }%
\providecommand \Eprint [0]{\href }%
\providecommand \doibase [0]{http://dx.doi.org/}%
\providecommand \selectlanguage [0]{\@gobble}%
\providecommand \bibinfo  [0]{\@secondoftwo}%
\providecommand \bibfield  [0]{\@secondoftwo}%
\providecommand \translation [1]{[#1]}%
\providecommand \BibitemOpen [0]{}%
\providecommand \bibitemStop [0]{}%
\providecommand \bibitemNoStop [0]{.\EOS\space}%
\providecommand \EOS [0]{\spacefactor3000\relax}%
\providecommand \BibitemShut  [1]{\csname bibitem#1\endcsname}%
\let\auto@bib@innerbib\@empty
\bibitem [{\citenamefont {Navr{\'{a}}til}\ \emph {et~al.}(2009)\citenamefont
  {Navr{\'{a}}til}, \citenamefont {Quaglioni}, \citenamefont {Stetcu},\ and\
  \citenamefont {Barrett}}]{NaQu09}%
  \BibitemOpen
  \bibfield  {author} {\bibinfo {author} {\bibfnamefont {P.}~\bibnamefont
  {Navr{\'{a}}til}}, \bibinfo {author} {\bibfnamefont {S.}~\bibnamefont
  {Quaglioni}}, \bibinfo {author} {\bibfnamefont {I.}~\bibnamefont {Stetcu}}, \
  and\ \bibinfo {author} {\bibfnamefont {B.}~\bibnamefont {Barrett}},\ }\href
  {\doibase 10.1088/0954-3899/36/8/083101} {\bibfield  {journal} {\bibinfo
  {journal} {J. Phys. G}\ }\textbf {\bibinfo {volume} {36}},\ \bibinfo {pages}
  {83101} (\bibinfo {year} {2009})}\BibitemShut {NoStop}%
\bibitem [{\citenamefont {Roth}\ \emph {et~al.}(2011)\citenamefont {Roth},
  \citenamefont {Langhammer}, \citenamefont {Calci}, \citenamefont {Binder},\
  and\ \citenamefont {Navr{\'{a}}til}}]{RoLa11}%
  \BibitemOpen
  \bibfield  {author} {\bibinfo {author} {\bibfnamefont {R.}~\bibnamefont
  {Roth}}, \bibinfo {author} {\bibfnamefont {J.}~\bibnamefont {Langhammer}},
  \bibinfo {author} {\bibfnamefont {A.}~\bibnamefont {Calci}}, \bibinfo
  {author} {\bibfnamefont {S.}~\bibnamefont {Binder}}, \ and\ \bibinfo {author}
  {\bibfnamefont {P.}~\bibnamefont {Navr{\'{a}}til}},\ }\href@noop {}
  {\bibfield  {journal} {\bibinfo  {journal} {Phys. Rev. Lett.}\ }\textbf
  {\bibinfo {volume} {107}},\ \bibinfo {pages} {72501} (\bibinfo {year}
  {2011})}\BibitemShut {NoStop}%
\bibitem [{\citenamefont {Barrett}\ \emph {et~al.}(2013)\citenamefont
  {Barrett}, \citenamefont {Navr{\'{a}}til},\ and\ \citenamefont
  {Vary}}]{BaNa13}%
  \BibitemOpen
  \bibfield  {author} {\bibinfo {author} {\bibfnamefont {B.~R.}\ \bibnamefont
  {Barrett}}, \bibinfo {author} {\bibfnamefont {P.}~\bibnamefont
  {Navr{\'{a}}til}}, \ and\ \bibinfo {author} {\bibfnamefont {J.~P.}\
  \bibnamefont {Vary}},\ }\href {\doibase 10.1016/j.ppnp.2012.10.003}
  {\bibfield  {journal} {\bibinfo  {journal} {Prog. Part. Nucl. Phys.}\
  }\textbf {\bibinfo {volume} {69}},\ \bibinfo {pages} {131} (\bibinfo {year}
  {2013})}\BibitemShut {NoStop}%
\bibitem [{\citenamefont {Gezerlis}\ \emph {et~al.}(2013)\citenamefont
  {Gezerlis}, \citenamefont {Tews}, \citenamefont {Epelbaum}, \citenamefont
  {Gandolfi}, \citenamefont {Hebeler}, \citenamefont {Nogga},\ and\
  \citenamefont {Schwenk}}]{Gezerlis2013}%
  \BibitemOpen
  \bibfield  {author} {\bibinfo {author} {\bibfnamefont {A.}~\bibnamefont
  {Gezerlis}}, \bibinfo {author} {\bibfnamefont {I.}~\bibnamefont {Tews}},
  \bibinfo {author} {\bibfnamefont {E.}~\bibnamefont {Epelbaum}}, \bibinfo
  {author} {\bibfnamefont {S.}~\bibnamefont {Gandolfi}}, \bibinfo {author}
  {\bibfnamefont {K.}~\bibnamefont {Hebeler}}, \bibinfo {author} {\bibfnamefont
  {A.}~\bibnamefont {Nogga}}, \ and\ \bibinfo {author} {\bibfnamefont
  {A.}~\bibnamefont {Schwenk}},\ }\href {\doibase
  10.1103/PhysRevLett.111.032501} {\bibfield  {journal} {\bibinfo  {journal}
  {Phys. Rev. Lett.}\ }\textbf {\bibinfo {volume} {111}},\ \bibinfo {pages}
  {032501} (\bibinfo {year} {2013})}\BibitemShut {NoStop}%
\bibitem [{\citenamefont {Carlson}\ \emph {et~al.}(2015)\citenamefont
  {Carlson}, \citenamefont {Gandolfi}, \citenamefont {Pederiva}, \citenamefont
  {Pieper}, \citenamefont {Schiavilla}, \citenamefont {Schmidt},\ and\
  \citenamefont {Wiringa}}]{Carlson:2015}%
  \BibitemOpen
  \bibfield  {author} {\bibinfo {author} {\bibfnamefont {J.}~\bibnamefont
  {Carlson}}, \bibinfo {author} {\bibfnamefont {S.}~\bibnamefont {Gandolfi}},
  \bibinfo {author} {\bibfnamefont {F.}~\bibnamefont {Pederiva}}, \bibinfo
  {author} {\bibfnamefont {S.~C.}\ \bibnamefont {Pieper}}, \bibinfo {author}
  {\bibfnamefont {R.}~\bibnamefont {Schiavilla}}, \bibinfo {author}
  {\bibfnamefont {K.~E.}\ \bibnamefont {Schmidt}}, \ and\ \bibinfo {author}
  {\bibfnamefont {R.~B.}\ \bibnamefont {Wiringa}},\ }\href {\doibase
  10.1103/RevModPhys.87.1067} {\bibfield  {journal} {\bibinfo  {journal} {Rev.
  Mod. Phys.}\ }\textbf {\bibinfo {volume} {87}},\ \bibinfo {pages} {1067}
  (\bibinfo {year} {2015})}\BibitemShut {NoStop}%
\bibitem [{\citenamefont {Lynn}\ \emph {et~al.}(2017)\citenamefont {Lynn},
  \citenamefont {Tews}, \citenamefont {Carlson}, \citenamefont {Gandolfi},
  \citenamefont {Gezerlis}, \citenamefont {Schmidt},\ and\ \citenamefont
  {Schwenk}}]{Lynn:2017}%
  \BibitemOpen
  \bibfield  {author} {\bibinfo {author} {\bibfnamefont {J.}~\bibnamefont
  {Lynn}}, \bibinfo {author} {\bibfnamefont {I.}~\bibnamefont {Tews}}, \bibinfo
  {author} {\bibfnamefont {J.}~\bibnamefont {Carlson}}, \bibinfo {author}
  {\bibfnamefont {S.}~\bibnamefont {Gandolfi}}, \bibinfo {author}
  {\bibfnamefont {A.}~\bibnamefont {Gezerlis}}, \bibinfo {author}
  {\bibfnamefont {K.}~\bibnamefont {Schmidt}}, \ and\ \bibinfo {author}
  {\bibfnamefont {A.}~\bibnamefont {Schwenk}},\ }\href@noop {} {\bibfield
  {journal} {\bibinfo  {journal} {Phys. Rev. C}\ }\textbf {\bibinfo {volume}
  {96}} (\bibinfo {year} {2017})}\BibitemShut {NoStop}%
\bibitem [{\citenamefont {Morris}\ \emph {et~al.}(2018)\citenamefont {Morris},
  \citenamefont {Simonis}, \citenamefont {Stroberg}, \citenamefont {Stumpf},
  \citenamefont {Hagen}, \citenamefont {Holt}, \citenamefont {Jansen},
  \citenamefont {Papenbrock}, \citenamefont {Roth},\ and\ \citenamefont
  {Schwenk}}]{Morris:2018}%
  \BibitemOpen
  \bibfield  {author} {\bibinfo {author} {\bibfnamefont {T.~D.}\ \bibnamefont
  {Morris}}, \bibinfo {author} {\bibfnamefont {J.}~\bibnamefont {Simonis}},
  \bibinfo {author} {\bibfnamefont {S.~R.}\ \bibnamefont {Stroberg}}, \bibinfo
  {author} {\bibfnamefont {C.}~\bibnamefont {Stumpf}}, \bibinfo {author}
  {\bibfnamefont {G.}~\bibnamefont {Hagen}}, \bibinfo {author} {\bibfnamefont
  {J.~D.}\ \bibnamefont {Holt}}, \bibinfo {author} {\bibfnamefont {G.~R.}\
  \bibnamefont {Jansen}}, \bibinfo {author} {\bibfnamefont {T.}~\bibnamefont
  {Papenbrock}}, \bibinfo {author} {\bibfnamefont {R.}~\bibnamefont {Roth}}, \
  and\ \bibinfo {author} {\bibfnamefont {A.}~\bibnamefont {Schwenk}},\ }\href
  {\doibase 10.1103/PhysRevLett.120.152503} {\bibfield  {journal} {\bibinfo
  {journal} {Phys. Rev. Lett.}\ }\textbf {\bibinfo {volume} {120}},\ \bibinfo
  {pages} {152503} (\bibinfo {year} {2018})}\BibitemShut {NoStop}%
\bibitem [{\citenamefont {Bogner}\ \emph {et~al.}(2005)\citenamefont {Bogner},
  \citenamefont {Schwenk}, \citenamefont {Furnstahl},\ and\ \citenamefont
  {Nogga}}]{Bogner:2005}%
  \BibitemOpen
  \bibfield  {author} {\bibinfo {author} {\bibfnamefont {S.}~\bibnamefont
  {Bogner}}, \bibinfo {author} {\bibfnamefont {A.}~\bibnamefont {Schwenk}},
  \bibinfo {author} {\bibfnamefont {R.}~\bibnamefont {Furnstahl}}, \ and\
  \bibinfo {author} {\bibfnamefont {A.}~\bibnamefont {Nogga}},\ }\href
  {\doibase 10.1016/j.nuclphysa.2005.08.024} {\bibfield  {journal} {\bibinfo
  {journal} {Nucl. Phys. A}\ }\textbf {\bibinfo {volume} {763}},\ \bibinfo
  {pages} {59} (\bibinfo {year} {2005})}\BibitemShut {NoStop}%
\bibitem [{\citenamefont {Roth}\ and\ \citenamefont
  {Langhammer}(2010)}]{Roth:2009up}%
  \BibitemOpen
  \bibfield  {author} {\bibinfo {author} {\bibfnamefont {R.}~\bibnamefont
  {Roth}}\ and\ \bibinfo {author} {\bibfnamefont {J.}~\bibnamefont
  {Langhammer}},\ }\href {\doibase 10.1016/j.physletb.2009.12.046} {\bibfield
  {journal} {\bibinfo  {journal} {Phys. Lett. B}\ }\textbf {\bibinfo {volume}
  {683}},\ \bibinfo {pages} {272} (\bibinfo {year} {2010})}\BibitemShut
  {NoStop}%
\bibitem [{\citenamefont {Hebeler}\ \emph {et~al.}(2011)\citenamefont
  {Hebeler}, \citenamefont {Bogner}, \citenamefont {Furnstahl}, \citenamefont
  {Nogga},\ and\ \citenamefont {Schwenk}}]{Hebeler:2011}%
  \BibitemOpen
  \bibfield  {author} {\bibinfo {author} {\bibfnamefont {K.}~\bibnamefont
  {Hebeler}}, \bibinfo {author} {\bibfnamefont {S.~K.}\ \bibnamefont {Bogner}},
  \bibinfo {author} {\bibfnamefont {R.~J.}\ \bibnamefont {Furnstahl}}, \bibinfo
  {author} {\bibfnamefont {A.}~\bibnamefont {Nogga}}, \ and\ \bibinfo {author}
  {\bibfnamefont {A.}~\bibnamefont {Schwenk}},\ }\href {\doibase
  10.1103/PhysRevC.83.031301} {\bibfield  {journal} {\bibinfo  {journal} {Phys.
  Rev. C}\ }\textbf {\bibinfo {volume} {83}},\ \bibinfo {pages} {031301}
  (\bibinfo {year} {2011})}\BibitemShut {NoStop}%
\bibitem [{\citenamefont {Langhammer}\ \emph {et~al.}(2012)\citenamefont
  {Langhammer}, \citenamefont {Roth},\ and\ \citenamefont
  {Stumpf}}]{Langhammer2012}%
  \BibitemOpen
  \bibfield  {author} {\bibinfo {author} {\bibfnamefont {J.}~\bibnamefont
  {Langhammer}}, \bibinfo {author} {\bibfnamefont {R.}~\bibnamefont {Roth}}, \
  and\ \bibinfo {author} {\bibfnamefont {C.}~\bibnamefont {Stumpf}},\ }\href
  {\doibase 10.1103/PhysRevC.86.054315} {\bibfield  {journal} {\bibinfo
  {journal} {Phys. Rev. C}\ }\textbf {\bibinfo {volume} {86}},\ \bibinfo
  {pages} {054315} (\bibinfo {year} {2012})}\BibitemShut {NoStop}%
\bibitem [{\citenamefont {Holt}\ \emph {et~al.}(2014)\citenamefont {Holt},
  \citenamefont {Men\'endez}, \citenamefont {Simonis},\ and\ \citenamefont
  {Schwenk}}]{Holt:2014}%
  \BibitemOpen
  \bibfield  {author} {\bibinfo {author} {\bibfnamefont {J.~D.}\ \bibnamefont
  {Holt}}, \bibinfo {author} {\bibfnamefont {J.}~\bibnamefont {Men\'endez}},
  \bibinfo {author} {\bibfnamefont {J.}~\bibnamefont {Simonis}}, \ and\
  \bibinfo {author} {\bibfnamefont {A.}~\bibnamefont {Schwenk}},\ }\href
  {\doibase 10.1103/PhysRevC.90.024312} {\bibfield  {journal} {\bibinfo
  {journal} {Phys. Rev. C}\ }\textbf {\bibinfo {volume} {90}},\ \bibinfo
  {pages} {024312} (\bibinfo {year} {2014})}\BibitemShut {NoStop}%
\bibitem [{\citenamefont {Tichai}\ \emph {et~al.}(2016)\citenamefont {Tichai},
  \citenamefont {Langhammer}, \citenamefont {Binder},\ and\ \citenamefont
  {Roth}}]{Tichai2016}%
  \BibitemOpen
  \bibfield  {author} {\bibinfo {author} {\bibfnamefont {A.}~\bibnamefont
  {Tichai}}, \bibinfo {author} {\bibfnamefont {J.}~\bibnamefont {Langhammer}},
  \bibinfo {author} {\bibfnamefont {S.}~\bibnamefont {Binder}}, \ and\ \bibinfo
  {author} {\bibfnamefont {R.}~\bibnamefont {Roth}},\ }\href {\doibase
  10.1016/j.physletb.2016.03.029} {\bibfield  {journal} {\bibinfo  {journal}
  {Phys. Lett. B}\ }\textbf {\bibinfo {volume} {756}},\ \bibinfo {pages} {283}
  (\bibinfo {year} {2016})}\BibitemShut {NoStop}%
\bibitem [{\citenamefont {Hu}\ \emph {et~al.}(2016)\citenamefont {Hu},
  \citenamefont {Xu}, \citenamefont {Sun}, \citenamefont {Vary},\ and\
  \citenamefont {Li}}]{Hu:2016}%
  \BibitemOpen
  \bibfield  {author} {\bibinfo {author} {\bibfnamefont {B.~S.}\ \bibnamefont
  {Hu}}, \bibinfo {author} {\bibfnamefont {F.~R.}\ \bibnamefont {Xu}}, \bibinfo
  {author} {\bibfnamefont {Z.~H.}\ \bibnamefont {Sun}}, \bibinfo {author}
  {\bibfnamefont {J.~P.}\ \bibnamefont {Vary}}, \ and\ \bibinfo {author}
  {\bibfnamefont {T.}~\bibnamefont {Li}},\ }\href {\doibase
  10.1103/PhysRevC.94.014303} {\bibfield  {journal} {\bibinfo  {journal} {Phys.
  Rev. C}\ }\textbf {\bibinfo {volume} {94}},\ \bibinfo {pages} {014303}
  (\bibinfo {year} {2016})}\BibitemShut {NoStop}%
\bibitem [{\citenamefont {Tichai}\ \emph
  {et~al.}(2018{\natexlab{a}})\citenamefont {Tichai}, \citenamefont
  {Gebrerufael}, \citenamefont {Vobig},\ and\ \citenamefont
  {Roth}}]{Tichai:2018ncsmpt}%
  \BibitemOpen
  \bibfield  {author} {\bibinfo {author} {\bibfnamefont {A.}~\bibnamefont
  {Tichai}}, \bibinfo {author} {\bibfnamefont {E.}~\bibnamefont {Gebrerufael}},
  \bibinfo {author} {\bibfnamefont {K.}~\bibnamefont {Vobig}}, \ and\ \bibinfo
  {author} {\bibfnamefont {R.}~\bibnamefont {Roth}},\ }\href {\doibase
  https://doi.org/10.1016/j.physletb.2018.10.029} {\bibfield  {journal}
  {\bibinfo  {journal} {Phys. Lett. B}\ }\textbf {\bibinfo {volume} {786}},\
  \bibinfo {pages} {448 } (\bibinfo {year} {2018}{\natexlab{a}})}\BibitemShut
  {NoStop}%
\bibitem [{\citenamefont {Tichai}\ \emph
  {et~al.}(2018{\natexlab{b}})\citenamefont {Tichai}, \citenamefont {Arthuis},
  \citenamefont {Duguet}, \citenamefont {Hergert}, \citenamefont {Somà},\ and\
  \citenamefont {Roth}}]{Tichai:2018mll}%
  \BibitemOpen
  \bibfield  {author} {\bibinfo {author} {\bibfnamefont {A.}~\bibnamefont
  {Tichai}}, \bibinfo {author} {\bibfnamefont {P.}~\bibnamefont {Arthuis}},
  \bibinfo {author} {\bibfnamefont {T.}~\bibnamefont {Duguet}}, \bibinfo
  {author} {\bibfnamefont {H.}~\bibnamefont {Hergert}}, \bibinfo {author}
  {\bibfnamefont {V.}~\bibnamefont {Somà}}, \ and\ \bibinfo {author}
  {\bibfnamefont {R.}~\bibnamefont {Roth}},\ }\href {\doibase
  https://doi.org/10.1016/j.physletb.2018.09.044} {\bibfield  {journal}
  {\bibinfo  {journal} {Phys. Lett. B}\ }\textbf {\bibinfo {volume} {786}},\
  \bibinfo {pages} {195 } (\bibinfo {year} {2018}{\natexlab{b}})}\BibitemShut
  {NoStop}%
\bibitem [{\citenamefont {Drischler}\ \emph {et~al.}(2019)\citenamefont
  {Drischler}, \citenamefont {Hebeler},\ and\ \citenamefont
  {Schwenk}}]{Drischler:2019}%
  \BibitemOpen
  \bibfield  {author} {\bibinfo {author} {\bibfnamefont {C.}~\bibnamefont
  {Drischler}}, \bibinfo {author} {\bibfnamefont {K.}~\bibnamefont {Hebeler}},
  \ and\ \bibinfo {author} {\bibfnamefont {A.}~\bibnamefont {Schwenk}},\ }\href
  {\doibase 10.1103/PhysRevLett.122.042501} {\bibfield  {journal} {\bibinfo
  {journal} {Phys. Rev. Lett.}\ }\textbf {\bibinfo {volume} {122}},\ \bibinfo
  {pages} {042501} (\bibinfo {year} {2019})}\BibitemShut {NoStop}%
\bibitem [{\citenamefont {Hagen}\ \emph {et~al.}(2010)\citenamefont {Hagen},
  \citenamefont {Papenbrock}, \citenamefont {Dean},\ and\ \citenamefont
  {Hjorth-Jensen}}]{HaPa10}%
  \BibitemOpen
  \bibfield  {author} {\bibinfo {author} {\bibfnamefont {G.}~\bibnamefont
  {Hagen}}, \bibinfo {author} {\bibfnamefont {T.}~\bibnamefont {Papenbrock}},
  \bibinfo {author} {\bibfnamefont {D.~J.}\ \bibnamefont {Dean}}, \ and\
  \bibinfo {author} {\bibfnamefont {M.}~\bibnamefont {Hjorth-Jensen}},\ }\href
  {\doibase 10.1103/PhysRevC.82.034330} {\bibfield  {journal} {\bibinfo
  {journal} {Phys. Rev. C}\ }\textbf {\bibinfo {volume} {82}},\ \bibinfo
  {pages} {34330} (\bibinfo {year} {2010})}\BibitemShut {NoStop}%
\bibitem [{\citenamefont {Binder}\ \emph {et~al.}(2014)\citenamefont {Binder},
  \citenamefont {Langhammer}, \citenamefont {Calci},\ and\ \citenamefont
  {Roth}}]{BiLa14}%
  \BibitemOpen
  \bibfield  {author} {\bibinfo {author} {\bibfnamefont {S.}~\bibnamefont
  {Binder}}, \bibinfo {author} {\bibfnamefont {J.}~\bibnamefont {Langhammer}},
  \bibinfo {author} {\bibfnamefont {A.}~\bibnamefont {Calci}}, \ and\ \bibinfo
  {author} {\bibfnamefont {R.}~\bibnamefont {Roth}},\ }\href@noop {} {\bibfield
   {journal} {\bibinfo  {journal} {Phys. Lett. B}\ }\textbf {\bibinfo {volume}
  {736}},\ \bibinfo {pages} {119} (\bibinfo {year} {2014})}\BibitemShut
  {NoStop}%
\bibitem [{\citenamefont {Henderson}\ \emph {et~al.}(2014)\citenamefont
  {Henderson}, \citenamefont {Dukelsky}, \citenamefont {Scuseria},
  \citenamefont {Signoracci},\ and\ \citenamefont
  {Duguet}}]{Henderson:2014vka}%
  \BibitemOpen
  \bibfield  {author} {\bibinfo {author} {\bibfnamefont {T.~M.}\ \bibnamefont
  {Henderson}}, \bibinfo {author} {\bibfnamefont {J.}~\bibnamefont {Dukelsky}},
  \bibinfo {author} {\bibfnamefont {G.~E.}\ \bibnamefont {Scuseria}}, \bibinfo
  {author} {\bibfnamefont {A.}~\bibnamefont {Signoracci}}, \ and\ \bibinfo
  {author} {\bibfnamefont {T.}~\bibnamefont {Duguet}},\ }\href@noop {}
  {\bibfield  {journal} {\bibinfo  {journal} {Phys. Rev. C}\ }\textbf {\bibinfo
  {volume} {89}},\ \bibinfo {pages} {054305} (\bibinfo {year}
  {2014})}\BibitemShut {NoStop}%
\bibitem [{\citenamefont {Hagen}\ \emph {et~al.}(2014)\citenamefont {Hagen},
  \citenamefont {Papenbrock}, \citenamefont {Hjorth-Jensen},\ and\
  \citenamefont {Dean}}]{Hagen:2014review}%
  \BibitemOpen
  \bibfield  {author} {\bibinfo {author} {\bibfnamefont {G.}~\bibnamefont
  {Hagen}}, \bibinfo {author} {\bibfnamefont {T.}~\bibnamefont {Papenbrock}},
  \bibinfo {author} {\bibfnamefont {M.}~\bibnamefont {Hjorth-Jensen}}, \ and\
  \bibinfo {author} {\bibfnamefont {D.~J.}\ \bibnamefont {Dean}},\ }\href
  {\doibase 10.1088/0034-4885/77/9/096302} {\bibfield  {journal} {\bibinfo
  {journal} {Rep. Prog. Phys.}\ }\textbf {\bibinfo {volume} {77}},\ \bibinfo
  {pages} {096302} (\bibinfo {year} {2014})}\BibitemShut {NoStop}%
\bibitem [{\citenamefont {Dickhoff}\ and\ \citenamefont
  {Barbieri}(2004)}]{Dickhoff:2004xx}%
  \BibitemOpen
  \bibfield  {author} {\bibinfo {author} {\bibfnamefont {W.~H.}\ \bibnamefont
  {Dickhoff}}\ and\ \bibinfo {author} {\bibfnamefont {C.}~\bibnamefont
  {Barbieri}},\ }\href@noop {} {\bibfield  {journal} {\bibinfo  {journal}
  {Prog. Part. Nucl. Phys.}\ }\textbf {\bibinfo {volume} {52}},\ \bibinfo
  {pages} {377} (\bibinfo {year} {2004})}\BibitemShut {NoStop}%
\bibitem [{\citenamefont {Cipollone}\ \emph {et~al.}(2013)\citenamefont
  {Cipollone}, \citenamefont {Barbieri},\ and\ \citenamefont
  {Navr{\'{a}}til}}]{CiBa13}%
  \BibitemOpen
  \bibfield  {author} {\bibinfo {author} {\bibfnamefont {A.}~\bibnamefont
  {Cipollone}}, \bibinfo {author} {\bibfnamefont {C.}~\bibnamefont {Barbieri}},
  \ and\ \bibinfo {author} {\bibfnamefont {P.}~\bibnamefont {Navr{\'{a}}til}},\
  }\href@noop {} {\bibfield  {journal} {\bibinfo  {journal} {Phys. Rev. Lett.}\
  }\textbf {\bibinfo {volume} {111}},\ \bibinfo {pages} {062501} (\bibinfo
  {year} {2013})}\BibitemShut {NoStop}%
\bibitem [{\citenamefont {Carbone}\ \emph {et~al.}(2013)\citenamefont
  {Carbone}, \citenamefont {Cipollone}, \citenamefont {Barbieri}, \citenamefont
  {Rios},\ and\ \citenamefont {Polls}}]{Carbone:2013eqa}%
  \BibitemOpen
  \bibfield  {author} {\bibinfo {author} {\bibfnamefont {A.}~\bibnamefont
  {Carbone}}, \bibinfo {author} {\bibfnamefont {A.}~\bibnamefont {Cipollone}},
  \bibinfo {author} {\bibfnamefont {C.}~\bibnamefont {Barbieri}}, \bibinfo
  {author} {\bibfnamefont {A.}~\bibnamefont {Rios}}, \ and\ \bibinfo {author}
  {\bibfnamefont {A.}~\bibnamefont {Polls}},\ }\href@noop {} {\bibfield
  {journal} {\bibinfo  {journal} {Phys. Rev. C}\ }\textbf {\bibinfo {volume}
  {88}},\ \bibinfo {pages} {54326} (\bibinfo {year} {2013})}\BibitemShut
  {NoStop}%
\bibitem [{\citenamefont {Som{\`{a}}}\ \emph {et~al.}(2014)\citenamefont
  {Som{\`{a}}}, \citenamefont {Cipollone}, \citenamefont {Barbieri},
  \citenamefont {Navr{\'{a}}til},\ and\ \citenamefont {Duguet}}]{SoCi13}%
  \BibitemOpen
  \bibfield  {author} {\bibinfo {author} {\bibfnamefont {V.}~\bibnamefont
  {Som{\`{a}}}}, \bibinfo {author} {\bibfnamefont {A.}~\bibnamefont
  {Cipollone}}, \bibinfo {author} {\bibfnamefont {C.}~\bibnamefont {Barbieri}},
  \bibinfo {author} {\bibfnamefont {P.}~\bibnamefont {Navr{\'{a}}til}}, \ and\
  \bibinfo {author} {\bibfnamefont {T.}~\bibnamefont {Duguet}},\ }\href@noop {}
  {\bibfield  {journal} {\bibinfo  {journal} {Phys. Rev. C}\ }\textbf {\bibinfo
  {volume} {89}},\ \bibinfo {pages} {61301} (\bibinfo {year}
  {2014})}\BibitemShut {NoStop}%
\bibitem [{\citenamefont {Raimondi}\ and\ \citenamefont
  {Barbieri}(2019)}]{Raimondi:2019}%
  \BibitemOpen
  \bibfield  {author} {\bibinfo {author} {\bibfnamefont {F.}~\bibnamefont
  {Raimondi}}\ and\ \bibinfo {author} {\bibfnamefont {C.}~\bibnamefont
  {Barbieri}},\ }\href@noop {} {\bibfield  {journal} {\bibinfo  {journal}
  {Phys. Rev. C}\ }\textbf {\bibinfo {volume} {99}},\ \bibinfo {pages} {054327}
  (\bibinfo {year} {2019})}\BibitemShut {NoStop}%
\bibitem [{\citenamefont {Tsukiyama}\ \emph {et~al.}(2011)\citenamefont
  {Tsukiyama}, \citenamefont {Bogner},\ and\ \citenamefont
  {Schwenk}}]{Tsukiyama:2011}%
  \BibitemOpen
  \bibfield  {author} {\bibinfo {author} {\bibfnamefont {K.}~\bibnamefont
  {Tsukiyama}}, \bibinfo {author} {\bibfnamefont {S.~K.}\ \bibnamefont
  {Bogner}}, \ and\ \bibinfo {author} {\bibfnamefont {A.}~\bibnamefont
  {Schwenk}},\ }\href {\doibase 10.1103/PhysRevLett.106.222502} {\bibfield
  {journal} {\bibinfo  {journal} {Phys. Rev. Lett.}\ }\textbf {\bibinfo
  {volume} {106}},\ \bibinfo {pages} {222502} (\bibinfo {year}
  {2011})}\BibitemShut {NoStop}%
\bibitem [{\citenamefont {Tsukiyama}\ \emph {et~al.}(2012)\citenamefont
  {Tsukiyama}, \citenamefont {Bogner},\ and\ \citenamefont
  {Schwenk}}]{Tsukiyama:2012}%
  \BibitemOpen
  \bibfield  {author} {\bibinfo {author} {\bibfnamefont {K.}~\bibnamefont
  {Tsukiyama}}, \bibinfo {author} {\bibfnamefont {S.~K.}\ \bibnamefont
  {Bogner}}, \ and\ \bibinfo {author} {\bibfnamefont {A.}~\bibnamefont
  {Schwenk}},\ }\href {https://link.aps.org/doi/10.1103/PhysRevC.85.061304}
  {\bibfield  {journal} {\bibinfo  {journal} {Phys. Rev. C}\ }\textbf {\bibinfo
  {volume} {85}},\ \bibinfo {pages} {061304} (\bibinfo {year}
  {2012})}\BibitemShut {NoStop}%
\bibitem [{\citenamefont {Hergert}\ \emph {et~al.}(2013)\citenamefont
  {Hergert}, \citenamefont {Bogner}, \citenamefont {Binder}, \citenamefont
  {Calci}, \citenamefont {Langhammer}, \citenamefont {Roth},\ and\
  \citenamefont {Schwenk}}]{HeBo13}%
  \BibitemOpen
  \bibfield  {author} {\bibinfo {author} {\bibfnamefont {H.}~\bibnamefont
  {Hergert}}, \bibinfo {author} {\bibfnamefont {S.~K.}\ \bibnamefont {Bogner}},
  \bibinfo {author} {\bibfnamefont {S.}~\bibnamefont {Binder}}, \bibinfo
  {author} {\bibfnamefont {A.}~\bibnamefont {Calci}}, \bibinfo {author}
  {\bibfnamefont {J.}~\bibnamefont {Langhammer}}, \bibinfo {author}
  {\bibfnamefont {R.}~\bibnamefont {Roth}}, \ and\ \bibinfo {author}
  {\bibfnamefont {A.}~\bibnamefont {Schwenk}},\ }\href@noop {} {\bibfield
  {journal} {\bibinfo  {journal} {Phys. Rev. C}\ }\textbf {\bibinfo {volume}
  {87}},\ \bibinfo {pages} {34307} (\bibinfo {year} {2013})}\BibitemShut
  {NoStop}%
\bibitem [{\citenamefont {Bogner}\ \emph {et~al.}(2014)\citenamefont {Bogner},
  \citenamefont {Hergert}, \citenamefont {Holt}, \citenamefont {Schwenk},
  \citenamefont {Binder}, \citenamefont {Calci}, \citenamefont {Langhammer},\
  and\ \citenamefont {Roth}}]{Bo14}%
  \BibitemOpen
  \bibfield  {author} {\bibinfo {author} {\bibfnamefont {S.~K.}\ \bibnamefont
  {Bogner}}, \bibinfo {author} {\bibfnamefont {H.}~\bibnamefont {Hergert}},
  \bibinfo {author} {\bibfnamefont {J.~D.}\ \bibnamefont {Holt}}, \bibinfo
  {author} {\bibfnamefont {A.}~\bibnamefont {Schwenk}}, \bibinfo {author}
  {\bibfnamefont {S.}~\bibnamefont {Binder}}, \bibinfo {author} {\bibfnamefont
  {A.}~\bibnamefont {Calci}}, \bibinfo {author} {\bibfnamefont
  {J.}~\bibnamefont {Langhammer}}, \ and\ \bibinfo {author} {\bibfnamefont
  {R.}~\bibnamefont {Roth}},\ }\href {\doibase 10.1103/PhysRevLett.113.142501}
  {\bibfield  {journal} {\bibinfo  {journal} {Phys. Rev. Lett.}\ }\textbf
  {\bibinfo {volume} {113}},\ \bibinfo {pages} {142501} (\bibinfo {year}
  {2014})}\BibitemShut {NoStop}%
\bibitem [{\citenamefont {Hergert}\ \emph {et~al.}(2016)\citenamefont
  {Hergert}, \citenamefont {Bogner}, \citenamefont {Morris}, \citenamefont
  {Schwenk},\ and\ \citenamefont {Tsukiyama}}]{H15}%
  \BibitemOpen
  \bibfield  {author} {\bibinfo {author} {\bibfnamefont {H.}~\bibnamefont
  {Hergert}}, \bibinfo {author} {\bibfnamefont {S.~K.}\ \bibnamefont {Bogner}},
  \bibinfo {author} {\bibfnamefont {T.~D.}\ \bibnamefont {Morris}}, \bibinfo
  {author} {\bibfnamefont {A.}~\bibnamefont {Schwenk}}, \ and\ \bibinfo
  {author} {\bibfnamefont {K.}~\bibnamefont {Tsukiyama}},\ }\href {\doibase
  10.1016/j.physrep.2015.12.007} {\bibfield  {journal} {\bibinfo  {journal}
  {Physics Reports}\ }\textbf {\bibinfo {volume} {621}},\ \bibinfo {pages}
  {165} (\bibinfo {year} {2016})}\BibitemShut {NoStop}%
\bibitem [{\citenamefont {Parzuchowski}\ \emph {et~al.}(2017)\citenamefont
  {Parzuchowski}, \citenamefont {Morris},\ and\ \citenamefont
  {Bogner}}]{Parzuchowski2017}%
  \BibitemOpen
  \bibfield  {author} {\bibinfo {author} {\bibfnamefont {N.~M.}\ \bibnamefont
  {Parzuchowski}}, \bibinfo {author} {\bibfnamefont {T.~D.}\ \bibnamefont
  {Morris}}, \ and\ \bibinfo {author} {\bibfnamefont {S.~K.}\ \bibnamefont
  {Bogner}},\ }\href {\doibase 10.1103/PhysRevC.95.044304} {\bibfield
  {journal} {\bibinfo  {journal} {Phys. Rev. C}\ }\textbf {\bibinfo {volume}
  {95}},\ \bibinfo {pages} {044304} (\bibinfo {year} {2017})}\BibitemShut
  {NoStop}%
\bibitem [{\citenamefont {Stroberg}\ \emph {et~al.}(2017)\citenamefont
  {Stroberg}, \citenamefont {Calci}, \citenamefont {Hergert}, \citenamefont
  {Holt}, \citenamefont {Bogner}, \citenamefont {Roth},\ and\ \citenamefont
  {Schwenk}}]{Stroberg2017}%
  \BibitemOpen
  \bibfield  {author} {\bibinfo {author} {\bibfnamefont {S.~R.}\ \bibnamefont
  {Stroberg}}, \bibinfo {author} {\bibfnamefont {A.}~\bibnamefont {Calci}},
  \bibinfo {author} {\bibfnamefont {H.}~\bibnamefont {Hergert}}, \bibinfo
  {author} {\bibfnamefont {J.~D.}\ \bibnamefont {Holt}}, \bibinfo {author}
  {\bibfnamefont {S.~K.}\ \bibnamefont {Bogner}}, \bibinfo {author}
  {\bibfnamefont {R.}~\bibnamefont {Roth}}, \ and\ \bibinfo {author}
  {\bibfnamefont {A.}~\bibnamefont {Schwenk}},\ }\href {\doibase
  10.1103/PhysRevLett.118.032502} {\bibfield  {journal} {\bibinfo  {journal}
  {Phys. Rev. Lett.}\ }\textbf {\bibinfo {volume} {118}},\ \bibinfo {pages}
  {032502} (\bibinfo {year} {2017})}\BibitemShut {NoStop}%
\bibitem [{\citenamefont {Kowalski}\ \emph {et~al.}(2004)\citenamefont
  {Kowalski}, \citenamefont {Dean}, \citenamefont {Hjorth-Jensen},
  \citenamefont {Papenbrock},\ and\ \citenamefont {Piecuch}}]{KoDe04}%
  \BibitemOpen
  \bibfield  {author} {\bibinfo {author} {\bibfnamefont {K.}~\bibnamefont
  {Kowalski}}, \bibinfo {author} {\bibfnamefont {D.~J.}\ \bibnamefont {Dean}},
  \bibinfo {author} {\bibfnamefont {M.}~\bibnamefont {Hjorth-Jensen}}, \bibinfo
  {author} {\bibfnamefont {T.}~\bibnamefont {Papenbrock}}, \ and\ \bibinfo
  {author} {\bibfnamefont {P.}~\bibnamefont {Piecuch}},\ }\href@noop {}
  {\bibfield  {journal} {\bibinfo  {journal} {Phys. Rev. Lett.}\ }\textbf
  {\bibinfo {volume} {92}},\ \bibinfo {pages} {132501} (\bibinfo {year}
  {2004})}\BibitemShut {NoStop}%
\bibitem [{\citenamefont {Barbieri}\ and\ \citenamefont
  {Dickhoff}(2002)}]{Barbieri2002}%
  \BibitemOpen
  \bibfield  {author} {\bibinfo {author} {\bibfnamefont {C.}~\bibnamefont
  {Barbieri}}\ and\ \bibinfo {author} {\bibfnamefont {W.~H.}\ \bibnamefont
  {Dickhoff}},\ }\href {\doibase 10.1103/PhysRevC.65.064313} {\bibfield
  {journal} {\bibinfo  {journal} {Phys. Rev. C}\ }\textbf {\bibinfo {volume}
  {65}},\ \bibinfo {pages} {064313} (\bibinfo {year} {2002})}\BibitemShut
  {NoStop}%
\bibitem [{\citenamefont {Bogner}\ \emph {et~al.}(2006)\citenamefont {Bogner},
  \citenamefont {Furnstahl}, \citenamefont {Ramanan},\ and\ \citenamefont
  {Schwenk}}]{Bogner:2006}%
  \BibitemOpen
  \bibfield  {author} {\bibinfo {author} {\bibfnamefont {S.}~\bibnamefont
  {Bogner}}, \bibinfo {author} {\bibfnamefont {R.}~\bibnamefont {Furnstahl}},
  \bibinfo {author} {\bibfnamefont {S.}~\bibnamefont {Ramanan}}, \ and\
  \bibinfo {author} {\bibfnamefont {A.}~\bibnamefont {Schwenk}},\ }\href
  {\doibase https://doi.org/10.1016/j.nuclphysa.2006.05.004} {\bibfield
  {journal} {\bibinfo  {journal} {Nucl. Phys. A}\ }\textbf {\bibinfo {volume}
  {773}},\ \bibinfo {pages} {203 } (\bibinfo {year} {2006})}\BibitemShut
  {NoStop}%
\bibitem [{\citenamefont {Ramanan}\ \emph {et~al.}(2007)\citenamefont
  {Ramanan}, \citenamefont {Bogner},\ and\ \citenamefont
  {Furnstahl}}]{Ramanan:2007}%
  \BibitemOpen
  \bibfield  {author} {\bibinfo {author} {\bibfnamefont {S.}~\bibnamefont
  {Ramanan}}, \bibinfo {author} {\bibfnamefont {S.}~\bibnamefont {Bogner}}, \
  and\ \bibinfo {author} {\bibfnamefont {R.}~\bibnamefont {Furnstahl}},\ }\href
  {\doibase https://doi.org/10.1016/j.nuclphysa.2007.10.005} {\bibfield
  {journal} {\bibinfo  {journal} {Nucl. Phys. A}\ }\textbf {\bibinfo {volume}
  {797}},\ \bibinfo {pages} {81} (\bibinfo {year} {2007})}\BibitemShut
  {NoStop}%
\bibitem [{\citenamefont {Hoppe}\ \emph {et~al.}(2017)\citenamefont {Hoppe},
  \citenamefont {Drischler}, \citenamefont {Furnstahl}, \citenamefont
  {Hebeler},\ and\ \citenamefont {Schwenk}}]{Hoppe:2017}%
  \BibitemOpen
  \bibfield  {author} {\bibinfo {author} {\bibfnamefont {J.}~\bibnamefont
  {Hoppe}}, \bibinfo {author} {\bibfnamefont {C.}~\bibnamefont {Drischler}},
  \bibinfo {author} {\bibfnamefont {R.~J.}\ \bibnamefont {Furnstahl}}, \bibinfo
  {author} {\bibfnamefont {K.}~\bibnamefont {Hebeler}}, \ and\ \bibinfo
  {author} {\bibfnamefont {A.}~\bibnamefont {Schwenk}},\ }\href {\doibase
  10.1103/PhysRevC.96.054002} {\bibfield  {journal} {\bibinfo  {journal} {Phys.
  Rev. C}\ }\textbf {\bibinfo {volume} {96}},\ \bibinfo {pages} {054002}
  (\bibinfo {year} {2017})}\BibitemShut {NoStop}%
\bibitem [{\citenamefont {Frame}\ \emph {et~al.}(2018)\citenamefont {Frame},
  \citenamefont {He}, \citenamefont {Ipsen}, \citenamefont {Lee}, \citenamefont
  {Lee},\ and\ \citenamefont {Rrapaj}}]{Frame:2018}%
  \BibitemOpen
  \bibfield  {author} {\bibinfo {author} {\bibfnamefont {D.}~\bibnamefont
  {Frame}}, \bibinfo {author} {\bibfnamefont {R.}~\bibnamefont {He}}, \bibinfo
  {author} {\bibfnamefont {I.}~\bibnamefont {Ipsen}}, \bibinfo {author}
  {\bibfnamefont {D.}~\bibnamefont {Lee}}, \bibinfo {author} {\bibfnamefont
  {D.}~\bibnamefont {Lee}}, \ and\ \bibinfo {author} {\bibfnamefont
  {E.}~\bibnamefont {Rrapaj}},\ }\href {\doibase
  10.1103/PhysRevLett.121.032501} {\bibfield  {journal} {\bibinfo  {journal}
  {Phys. Rev. Lett.}\ }\textbf {\bibinfo {volume} {121}},\ \bibinfo {pages}
  {032501} (\bibinfo {year} {2018})}\BibitemShut {NoStop}%
\bibitem [{\citenamefont {K\"onig}\ \emph {et~al.}(2019)\citenamefont
  {K\"onig}, \citenamefont {Ekstr{\"o}m}, \citenamefont {Hebeler},
  \citenamefont {Lee},\ and\ \citenamefont {Schwenk}}]{Koenig:2019}%
  \BibitemOpen
  \bibfield  {author} {\bibinfo {author} {\bibfnamefont {S.}~\bibnamefont
  {K\"onig}}, \bibinfo {author} {\bibfnamefont {A.}~\bibnamefont
  {Ekstr{\"o}m}}, \bibinfo {author} {\bibfnamefont {K.}~\bibnamefont
  {Hebeler}}, \bibinfo {author} {\bibfnamefont {D.}~\bibnamefont {Lee}}, \ and\
  \bibinfo {author} {\bibfnamefont {A.}~\bibnamefont {Schwenk}},\ }\href@noop
  {} {\  (\bibinfo {year} {2019})},\ \Eprint {http://arxiv.org/abs/1909.08446}
  {arXiv:1909.08446} \BibitemShut {NoStop}%
\bibitem [{\citenamefont {Sarkar}\ and\ \citenamefont
  {Lee}(2019)}]{Sarkar:2019}%
  \BibitemOpen
  \bibfield  {author} {\bibinfo {author} {\bibfnamefont {A.}~\bibnamefont
  {Sarkar}}\ and\ \bibinfo {author} {\bibfnamefont {D.}~\bibnamefont {Lee}},\
  }\href@noop {} {} (\bibinfo {year} {2019}),\ \bibinfo {note}
  {(unpublished)}\BibitemShut {NoStop}%
\bibitem [{\citenamefont {Shavitt}\ and\ \citenamefont
  {Bartlett}(2009)}]{Shavitt2009}%
  \BibitemOpen
  \bibfield  {author} {\bibinfo {author} {\bibfnamefont {I.}~\bibnamefont
  {Shavitt}}\ and\ \bibinfo {author} {\bibfnamefont {R.~J.}\ \bibnamefont
  {Bartlett}},\ }\href
  {http://www.amazon.com/Many-Body-Methods-Chemistry-Physics-Coupled-Cluster/dp/052181832X}
  {\emph {\bibinfo {title} {{Many-Body Methods in Chemistry and Physics: MBPT
  and Coupled-Cluster Theory (Cambridge Molecular Science)}}}}\ (\bibinfo
  {publisher} {Cambridge University Press},\ \bibinfo {year}
  {2009})\BibitemShut {NoStop}%
\bibitem [{\citenamefont {Entem}\ and\ \citenamefont
  {Machleidt}(2003)}]{Entem2003a}%
  \BibitemOpen
  \bibfield  {author} {\bibinfo {author} {\bibfnamefont {D.~R.}\ \bibnamefont
  {Entem}}\ and\ \bibinfo {author} {\bibfnamefont {R.}~\bibnamefont
  {Machleidt}},\ }\href {\doibase 10.1103/PhysRevC.68.041001} {\bibfield
  {journal} {\bibinfo  {journal} {Phys. Rev. C}\ }\textbf {\bibinfo {volume}
  {68}},\ \bibinfo {pages} {041001} (\bibinfo {year} {2003})}\BibitemShut
  {NoStop}%
\bibitem [{\citenamefont {Bogner}\ \emph {et~al.}(2007)\citenamefont {Bogner},
  \citenamefont {Furnstahl},\ and\ \citenamefont {Perry}}]{Bogner:2007}%
  \BibitemOpen
  \bibfield  {author} {\bibinfo {author} {\bibfnamefont {S.~K.}\ \bibnamefont
  {Bogner}}, \bibinfo {author} {\bibfnamefont {R.~J.}\ \bibnamefont
  {Furnstahl}}, \ and\ \bibinfo {author} {\bibfnamefont {R.~J.}\ \bibnamefont
  {Perry}},\ }\href {\doibase 10.1103/PhysRevC.75.061001} {\bibfield  {journal}
  {\bibinfo  {journal} {Phys. Rev. C}\ }\textbf {\bibinfo {volume} {75}},\
  \bibinfo {pages} {061001} (\bibinfo {year} {2007})}\BibitemShut {NoStop}%
\bibitem [{\citenamefont {Navratil}\ \emph {et~al.}(2000)\citenamefont
  {Navratil}, \citenamefont {Kamuntavicius},\ and\ \citenamefont
  {Barrett}}]{Navratil:1999pw}%
  \BibitemOpen
  \bibfield  {author} {\bibinfo {author} {\bibfnamefont {P.}~\bibnamefont
  {Navratil}}, \bibinfo {author} {\bibfnamefont {G.~P.}\ \bibnamefont
  {Kamuntavicius}}, \ and\ \bibinfo {author} {\bibfnamefont {B.~R.}\
  \bibnamefont {Barrett}},\ }\href {\doibase 10.1103/PhysRevC.61.044001}
  {\bibfield  {journal} {\bibinfo  {journal} {Phys. Rev. C}\ }\textbf {\bibinfo
  {volume} {61}},\ \bibinfo {pages} {044001} (\bibinfo {year}
  {2000})}\BibitemShut {NoStop}%
\bibitem [{\citenamefont {{Baker, Jr}.}\ and\ \citenamefont
  {Graves-Morris}(1996)}]{Baker:1996}%
  \BibitemOpen
  \bibfield  {author} {\bibinfo {author} {\bibfnamefont {G.~A.}\ \bibnamefont
  {{Baker, Jr}.}}\ and\ \bibinfo {author} {\bibfnamefont {P.}~\bibnamefont
  {Graves-Morris}},\ }\href@noop {} {\emph {\bibinfo {title} {Pad{\'e}
  approximants}}}\ (\bibinfo  {publisher} {Cambridge University Press},\
  \bibinfo {year} {1996})\BibitemShut {NoStop}%
\bibitem [{\citenamefont {Duguet}\ and\ \citenamefont
  {Signoracci}(2017)}]{Duguet:2015yle}%
  \BibitemOpen
  \bibfield  {author} {\bibinfo {author} {\bibfnamefont {T.}~\bibnamefont
  {Duguet}}\ and\ \bibinfo {author} {\bibfnamefont {A.}~\bibnamefont
  {Signoracci}},\ }\href@noop {} {\bibfield  {journal} {\bibinfo  {journal} {J.
  Phys.}\ }\textbf {\bibinfo {volume} {G44}},\ \bibinfo {pages} {15103}
  (\bibinfo {year} {2017})}\BibitemShut {NoStop}%
\bibitem [{\citenamefont {Demol}\ \emph {et~al.}(2019)\citenamefont {Demol},
  \citenamefont {Frosini}, \citenamefont {Tichai}, \citenamefont {Ripoche},
  \citenamefont {Som\`a},\ and\ \citenamefont {Duguet}}]{demol19a}%
  \BibitemOpen
  \bibfield  {author} {\bibinfo {author} {\bibfnamefont {P.}~\bibnamefont
  {Demol}}, \bibinfo {author} {\bibfnamefont {M.}~\bibnamefont {Frosini}},
  \bibinfo {author} {\bibfnamefont {A.}~\bibnamefont {Tichai}}, \bibinfo
  {author} {\bibfnamefont {J.}~\bibnamefont {Ripoche}}, \bibinfo {author}
  {\bibfnamefont {V.}~\bibnamefont {Som\`a}}, \ and\ \bibinfo {author}
  {\bibfnamefont {T.}~\bibnamefont {Duguet}},\ }\href@noop {} {} (\bibinfo
  {year} {2019}),\ \bibinfo {note} {(unpublished)}\BibitemShut {NoStop}%
\bibitem [{\citenamefont {Ring}\ and\ \citenamefont {Schuck}(1980)}]{RiSc80}%
  \BibitemOpen
  \bibfield  {author} {\bibinfo {author} {\bibfnamefont {P.}~\bibnamefont
  {Ring}}\ and\ \bibinfo {author} {\bibfnamefont {P.}~\bibnamefont {Schuck}},\
  }\href@noop {} {\emph {\bibinfo {title} {{The Nuclear Many-Body Problem}}}}\
  (\bibinfo  {publisher} {Springer Verlag, New York},\ \bibinfo {year}
  {1980})\BibitemShut {NoStop}%
\bibitem [{\citenamefont {Tichai}\ \emph {et~al.}(2019)\citenamefont {Tichai},
  \citenamefont {Ripoche},\ and\ \citenamefont {Duguet}}]{Tichai:2019ksh}%
  \BibitemOpen
  \bibfield  {author} {\bibinfo {author} {\bibfnamefont {A.}~\bibnamefont
  {Tichai}}, \bibinfo {author} {\bibfnamefont {J.}~\bibnamefont {Ripoche}}, \
  and\ \bibinfo {author} {\bibfnamefont {T.}~\bibnamefont {Duguet}},\
  }\href@noop {} {\bibfield  {journal} {\bibinfo  {journal} {Eur. Phys. J. A}\
  }\textbf {\bibinfo {volume} {55}} (\bibinfo {year} {2019})}\BibitemShut
  {NoStop}%
\bibitem [{\citenamefont {Ekstr{\"o}m}\ and\ \citenamefont
  {Hagen}(2019)}]{Ekstroem:2019}%
  \BibitemOpen
  \bibfield  {author} {\bibinfo {author} {\bibfnamefont {A.}~\bibnamefont
  {Ekstr{\"o}m}}\ and\ \bibinfo {author} {\bibfnamefont {G.}~\bibnamefont
  {Hagen}},\ }\href@noop {} {\  (\bibinfo {year} {2019})},\ \Eprint
  {http://arxiv.org/abs/1910.02922} {arXiv:1910.02922} \BibitemShut {NoStop}%
\end{thebibliography}%
